\documentclass[12pt,a4paper]{article}
\usepackage{a4wide}
\usepackage{latexsym}
\usepackage{amssymb}
\usepackage{amsmath, cite}
\usepackage{color}

 \makeatletter
\@addtoreset{equation}{section} \makeatother

\def\bfone{\relax{\rm 1\kern-.35em 1}}

 \makeatletter
\@addtoreset{equation}{section} \makeatother

\renewcommand{\l}{\langle}

\def\be {\begin{equation}}
\def\ee {\end{equation}}
\def\bea {\begin{eqnarray}}
\def\eea {\end{eqnarray}}
\def\bc {\begin{center}}
\def\ec {\end{center}}
\def\a  {\alpha}
\def\b  {\beta}

\def\D  {\Delta}

\def\l  {\lambda}

\def\m  {\mu}

\def\O  {\Omega}
\def\p  {\pi}

\def\t  {\tau}

\def\bfg {\begin{figure}}
\def\efg {\end{figure}}
\def\bi {\begin{itemize}}
\def\ei {\end{itemize}}
\def\nn {\nonumber}

\def\la {\label}
\def\le {\left}
\def\ri {\right}

\def\fr {\frac}

\DeclareFontFamily{U}{rsf}{} \DeclareFontShape{U}{rsf}{m}{n}{
  <5> <6> rsfs5 <7> <8> <9> rsfs7 <10-> rsfs10}{}
\DeclareMathAlphabet\Scr{U}{rsf}{m}{n}

\begin{document}

\begin{center}
{\bf \large{ No Existence of Black Holes at LHC Due to Minimal Length in Quantum Gravity \\[10mm]}}
\large {Ahmed Farag Ali} \\[4mm]
\small{Dept. of Physics, Faculty of Sciences, Benha
University, Benha 13518, Egypt.}\\[4mm]
{\upshape{ahmed.ali@fsc.bu.edu.eg ; ahmed.ali@uleth.ca}}\\[7mm]
\end{center}

\begin{abstract}
We investigate the impact of the Generalized Uncertainty
Principle (GUP), proposed by some approaches to quantum gravity
such as String Theory and Doubly Special Relativity Theories
(DSR) on the production of mini black holes, and show that the
minimum black hole mass is formed at energies higher than
the energy scales of LHC which possibly agrees with the recent
experimental results of LHC \cite{EXPERIMENT,EXPERIMENT2}
\end{abstract}

\section{Introduction}

The existence of a minimal length is one of the most
interesting predictions of some approaches related to quantum
gravity such as String Theory as well as Black hole physics.
This is a consequence of String Theory since strings can not
interact at distances smaller than their size, which yields {\it
Generalized Uncertainty Principle} (GUP) \cite{guppapers}. From
Black hole physics, the  Uncertainty Principle,
$\Delta x \sim \hbar/\Delta p$, is modified at
the Planck energy scale, when the corresponding Schwarzschild
radius is comparable to the Compton wavelength (both are
approximately equal to the Planck length). Higher energies
result in a further increase of the Schwarzschild radius,
resulting in $\Delta x \approx \ell_{Pl}^2\Delta p/\hbar$ The
above observation, along with a combination of thought
experiments and rigorous derivations suggest that the {\it
Generalized Uncertainty Principle} (GUP) holds at all scales,
and is represented by \cite{guppapers,BHGUP,kmm,kempf,brau,Hossenfelder:2012jw}

%
\bea \Delta x_i \Delta p_i &\geq& \fr{\hbar}{2} [ 1 + \beta
\le((\Delta p)^2 + <p>^2 \ri) \nn \\
&+& 2\beta \le( \Delta p_i^2 + <p_i>^2\ri) ]~, \la{uncert1}
\eea

where $p^2 = \sum\limits_{j}p_{j}p_{j}$,
$\beta=\beta_0/(M_{p}c)^2=\b_0 \frac{\ell_{p}^2}{\hbar^2}$, $M_{p}=$
Planck mass, and $M_{p} c^2=$ Planck energy.

It was shown in \cite{kmm}, that inequality (\ref{uncert1}) is
equivalent to the following modified Heisenberg algebra
\be [x_i,p_j] = i \hbar ( \delta_{ij} + \beta \delta_{ij} p^2 +
2\beta p_i p_j )~. \la{com1} \ee
This form ensures, via the Jacobi identity, that
$[x_i,x_j]=0=[p_i,p_j]$ \cite{kempf}.\\


Recently, we proposed the GUP in
\cite{advplb,Ali:2010yn,Das:2010zf} which predicts maximum
observable momenta  besides the existence of minimal measurable
length and is consistent with  Doubly Special Relativity (DSR)
theories, String Theory and Black Holes Physics {\it and} which
ensures $[x_i,x_j]=0=[p_i,p_j]$ (via the Jacobi identity).

\bea [x_i, p_j]\hspace{-1ex} &=&\hspace{-1ex} i
\hbar\hspace{-0.5ex} \left[  \delta_{ij}\hspace{-0.5ex} -
\hspace{-0.5ex} \alpha \hspace{-0.5ex}  \le( p \delta_{ij} +
\frac{p_i p_j}{p} \ri)
+ \alpha^2 \hspace{-0.5ex}
\le( p^2 \delta_{ij}  + 3 p_{i} p_{j} \ri) \hspace{-0.5ex} \ri]
\label{comm01}
\eea
where
$\alpha = {\alpha_0}/{M_{p}c} = {\alpha_0 \ell_{p}}/{\hbar},$
$M_{p}=$ Planck mass, $\ell_{p}=$ Planck length,
and $M_{p} c^2=$ Planck energy.

Note that  Eq. (\ref{comm01}) is approximately covariant under
DSR transformations \cite{cg}. Since DSR transformations
preserve both speed of light, and  invariant energy scale
\cite{sm}, it is not surprising that Eqs. (\ref{comm01}) imply
the existence of minimum measurable length and maximum
measurable momentum
\bea
\D x &\geq& (\D x)_{min}  \approx \alpha_0\ell_{p} \la{dxmin} \\
\D p &\leq& (\D p)_{max} \approx \frac{M_{p}c}{\alpha_0}~. \la{dpmax}
\eea
\par\noindent

Our proposed GUP suggests that the space is discrete \cite{advplb,Ali:2010yn,Das:2010zf} and that
all measurable lengths are quantized in units of a fundamental
minimum measurable length (which can be the Planck length).
Note that similar quantization of length was shown in the
context of Loop Quantum Gravity in \cite{LQG}.

Since the GUP modifies the fundamental commutator bracket
between position and momentum, naturally it modifies the
Hamiltonian and hence it affects a host of quantum phenomena,
and it is important to make a quantitative study of these
effects which would open a window for quantum gravity phenomenology.
In a series of papers, the author with collaborators
investigated the effects of GUP on atomic, condensed matter
systems and preheating phase of the universe
\cite{dvprl,Ali:2010yn,dvcjp, Ali:2011fa,Chemissany:2011nq}.
Also the author studied in\cite{faragali} its effects on the weak equivalence principle
(WEP) and the Liouville theorem (LT) in statistical mechanics,
and it was found that the GUP can potentially explain the small
observed violations of the WEP in neutron interferometry
experiments\cite{exp} and also predicts a modified invariant
phase space which is relevant to the Liouville theorem. Recently, it was
suggested in \cite{Pikovski:2011zk} that the GUP can be measured
directly in Quantum Optics Lab which confirm the theoretical
predictions in \cite{Ali:2011fa,dvprl}

The proposals for the existence of extra dimensions has opened
up new  doors of research in quantum gravity
\cite{ArkaniHamed:1998rs,Antoniadis:1998ig,Randall:1999ee,Randall:1999vf}.
In particular, a host of interesting work is being done on
different aspects of low-energy quantum gravity
phenomenology. One of the most significant sub-fields is the
study of black hole (BH) and brane production at the
LHC\cite{Dimopoulos:2001hw}.

In this paper, we present a phenomenological study of black
holes in higher dimensions at the Large Hadron Collider
(LHC) if GUP that follows from Jacobi Identity is taken into
consideration, see Eqs.\ (\ref{uncert1}, \ref{comm01}). If the
black hole can be produced and detected, it would result in an
additional mass threshold above the Planck scale at which new
physics can be found. The scope of the present work is to
investigate the effect of GUP on the Hawking temperature,
entropy, and BH decay rate.  We find that the BH
thermodynamics dramatically changed if the GUP parameter is
non-vanishing.

We also obtain an interesting result that black holes may not
be detectable at the current LHC energy scales. This result
possibly agrees with the recent experiments that were done
at LHC \cite{EXPERIMENT,EXPERIMENT2}, which tend to exclude
the black hole observations in the current energy scales at LHC.
The effect of the GUP on Black holes has been studied before
with different versions of GUP which does not follow from Jacobi Identity,
see e.g. \cite{Cavaglia:2003qk}, however the previous studies
predicted that BH's can be seen at the LHC
energy scales in disagreement with  the recent experimental
results of LHC\cite{EXPERIMENT,EXPERIMENT2}.
The suppressed black hole production has been studied with
with different approaches such as the non commutative geometry
and minimal length theories\cite{SuppressedBH}.

An outline of this paper is as follows. In Sec .2, We introduce
briefly how standard Hawking temperature can be obtained from
the standard uncertainty principle in $D$- dimensions. In Sec
.3, we investigate BH thermodynamics if GUP of Eq.\
(\ref{uncert1}) which was proposed earlier by
\cite{guppapers,BHGUP,kmm,kempf,brau} is taken into
consideration. We found that the BH minimal mass could be
detected in the range between  $60$ TeV and $2.3 \times
10^5$ TeV depending on the number of extra dimensions for
$D=6..10$. In Sec. 4, we repeat the analysis for  our proposed
model of GUP  in Eq.\ (\ref{comm01}). We find that BH's minimal
masses could be found in the energy scales between $13.8$ TeV
and $5.5 \times 10^3$ TeV for $D=6..10$. The results are explained in
Sec. 5. So, the threshold is less for the second case.

\section{Hawking Temperature -- Uncertainty Relation Connection}

In this section, we review the connection between standard
Hawking temperature and uncertainty relation that has been
proposed by Adler et al. in \cite{Adler}  and has been
generalized in large extra dimensions by  Cavaglia et al in
\cite{Cavaglia:2003qk}. A BH  could be modeled as
$(D-1)$-dimensional sphere  of size equal to twice of
Schwarzschild radius, $r_s$. Since the Hawking radiation is a
quantum process, so the emitted particle should obey the
Heisenberg uncertainty relation. This leads to
momentum-position uncertainty,

\be \Delta p_i \Delta x_j \geq \frac{\hbar}{2}
\delta_{ij}\label{SUP} , \ee

where the uncertainty in position of emitted Hawking particle
has its minimum value given by

\be \Delta x \approx 2 r_s =2 \lambda_D
\le[\frac{G_DM}{c^2}\ri]^{\frac{1}{D-3}}\label{DU}, \ee

where $\l_D=\le[\frac{16\p}{(D-2)\O_{D-2}}\ri]^{\frac{1}{D-3}}$
and $\O_D=\frac{2 \pi^{\frac{D-1}{2}}}{\Gamma(\frac{D-1}{2})}$.

Using Eq (\ref{SUP},\ref{DU}) with the argument used in
 \cite{Cavaglia:2003qk} that $\Delta x_i \Delta p_i \approx
 \Delta x \Delta p$ , the energy uncertainty of the emitted
 Hawking
particle  is given by

\be \D E\approx c~\D p=  c~ \frac{\hbar}{2~ \D x} \approx c~
\frac{\hbar}{4~ r_s}
=\frac{M_{p}c^2}{4\l_D}\le(\frac{M}{M_{p}}\ri)^{\frac{-1}{(D-3)}}\,.
\label{EU} \ee

From now on, we can assume $m=\frac{M}{M_p}$, where $m$ is the
mass in units of the Planck mass and the Planck mass $M_p$ is
given by $M_p=
[\frac{\hbar^{D-3}}{c^{D-5}G_D}]^{\frac{1}{(D-2)}}$ in
D-dimensions. 
As proposed by Adler et al. in \cite{Adler}, one can identify
the energy uncertainty $\D E$ as the energy of the emitted
photon from the black hole. Based on this argument, one can get
the characteristic temperature of the emitted Hawking particle
from the previous energy  by just multiplying it with a
calibration factor $\le(\frac{D-3}{\pi}\ri)$ \cite{Adler,Cavaglia:2003qk}to give exactly
the Hawking temperature \cite{Hawking:1974sw} in D-dimensions
of the spacetime as follows:

\be
T_H=\frac{D-3}{4\p\l_D}~M_{p}c^2~m^{\frac{-1}{(D-3)}}\,.
\la{hawT}
\ee

The thermodynamical properties of the BH can be computed via
the usual thermodynamic relations. The entropy can be
calculated using the first law of black hole thermodynamics,

\be
dM=\frac{1}{c^2} T d S \, .
\label{TD}
\ee

Using the mass in units of the Planck mass, $m$,  one can
rewrite Eq.\ (\ref{TD}) as follows: \be dS= M_p c^2 \frac{1}{T}
dm\; .\la{TD1} \ee By integrating Eq.\ (\ref{TD1}) using Eq.\
(\ref{hawT}), one can obtain the the Bekenstein
entropy\cite{Bekenstein:1973ur} as follows

\be
S=\frac{4\pi\l_D}{D-2}m^{(D-2)/(D-3)}\, .\la{entropy}
\ee

The specific heat can be calculated using the thermodynamical
relation

\be {\cal C}= T \frac{\partial S}{\partial T}= T \frac{\partial
S}{\partial m}\frac{\partial m}{\partial T}=
 M_p c^2 \frac{\partial m}{\partial T}\,,\la{SPH}
\ee
where we have used Eq.\ (\ref{TD1}) in the last equation.

By differentiating Eq.\ (\ref{hawT}) and substituting this into
Eq.\ (\ref{SPH})  , the specific heat could be given by \be
{\cal C} = -4\pi\l_D m^{\frac{(D-2)}{(D-3)}}\,, \la{C0} \ee

The Hawking temperature $T_H$ can be used in the calculation of
the emission rate. The emission rate might be calculated using
Stefan-Botlzmann law if the energy loss was dominated by
photons. Assuming $\cal D$-dimensional spacetime brane, the
thermal emission in the bulk of the brane can be neglected and
the black hole is supposed to radiate mainly on the brane
\cite{evaporation}, so the emission rate on the brane can be
given by:

\be
\frac{dM}{dt} \propto T^{{\cal D}}\,,
\label{dMdt}
\ee

Because the the black hole radiates mainly on the brane
\cite{evaporation},  i.e ${\cal{D}}=4$, the emission rate can
be found as following:

\be
\frac{dm}{dt}=-\frac{\m^{\prime}}{t_{p}}~m^{\frac{-2}{(D-3)}}\,,
\la{rate1}
\ee
where $t_{p}=\le(\frac{\hbar
G_D}{c^{D+1}}\ri)^{\frac{1}{(D-2)}}$ is the Planck time,
and the form of $\mu^{\prime}$ can be found in \cite{Cavaglia:2003qk}.\\

The decay time of the black hole can be obtained by integrating
Eq.\ (\ref{rate1}) to give

\be \t =  \m^{\prime -1} \le(\frac{D-3}{D-1}
\ri)m_i^{\frac{(D-1)}{(D-3)}}~t_{p}\,, \la{decayt} \ee Note
that the calculated Hawking temperature $T_H$, Bekenstein
entropy $S$, specific heat ${\cal C}$, emission rate
$\frac{dm}{dt}$, and decay time $\t$ lead to
\emph{\textbf{catastrophic evaporation} }as $m \rightarrow 0$.
This can be explained as following. Since ${\cal C}=0$ only
when $m=0$, the black hole will continue to radiate until
$m=0$. But as the black hole approaches zero mass, its
temperature approaches infinity with infinite radiation rate.
This was just a brief summary for the Hawking
radiation-Uncertainty principle connection, and the
catastrophic implications of Hawking radiation as the black
hole mass approaches zero. In the next two sections, we study
BH thermodynamics if GUP is taken into consideration. The
end-point of Hawking radiation is not catastrophic because GUP
imply the existence of BH remnants at which the specific heat
vanishes and, therefore, the BH cannot exchange heat with the
surrounding space. The GUP prevents BHs from evaporating
completely, just like the standard uncertainty principle
prevents the hydrogen atom from
collapsing\cite{Adler,Cavaglia:2003qk}.

\section{GUP quadratic in $\D p$ and BH thermodynamics }

In this section, we make analysis of BH thermodynamics
If  GUP proposed in \cite{guppapers,BHGUP,kmm,kempf,brau}
is taken into consideration.

%
%


The emitted particles as Hawking radiation are mostly photons
and standard model (SM) particles. According to the ADD model
of extra dimensions \cite{ArkaniHamed:1998rs}, photons and SM
particles are localized to the brane. So the photons or SM
particles have mainly $4$-components momentum and the other
components in the extra dimensions are equal to zero. For
simplicity, we might assume from kinetic theory of gases which
assumes a cloud of points in velocity space, equally spread in
all directions (there is no reason particle would prefer to be
moving in the x-direction, say, rather than the y-direction)
and consider:

\be
p_1\approx p_2\approx p_3\la{sym}
\ee

This assumption leads to

\bea
p^2&=& \sum_{i=1}^{3} p_i p_i \approx 3~ p_i^2\, \nn\\
\langle p_i^2 \rangle &\approx& \frac{1}{3}~ \langle p^2 \rangle\,.\la{3n}
\eea

So Eq.\ (\ref{uncert1}) reads, with using the argument used in
\cite{Cavaglia:2003qk},

\be \Delta x \Delta p \geq \frac{\hbar}{2} \le[ 1 +
\frac{5}{3}~ \b ~\langle p^2 \rangle\ri], \la{uncert3} \ee

%

Now, we want to find the relation between $\langle p^2 \rangle$
and $ \D p^2$. We can assume that we have a photon gas emitted
from the BH like emission from a black body. Therefore, we
might use Wien's  Law which gives  a temperature corresponding to
a peak emission at energy given by

\bea c~\langle p \rangle~= ~2.821~~ T_H, \eea From
Hawking-Uncertainty connection proposed by Adler et al. in
\cite{Adler}  and that was generalized in large extra
dimensions by  Cavaglia et al in \cite{Cavaglia:2003qk}, we
have \be T_H= \frac{D-3}{\pi}~ c ~\D p = \frac{1}{2.821}~ c~~
\langle p \rangle~. \ee

We get the following relations using  the relation $\langle p^2
\rangle = \D p^2 + {\langle p \rangle}^2$

\bea
\langle p \rangle ~&=& 2.821~ \frac{D-3}{\pi}~ \D p~ =~ \sqrt{\mu}~~ \D p \nn,\\
\langle p^2 \rangle~ &=& ~(1+\mu) ~\D p^2, ~~~~~\mbox{
where}~~~~~ \mu= \le(2.821~ \frac{D-3}{\pi}\ri)^2. \la{arg}
\eea


Using Eqs.\ (\ref{3n},\ref{arg}) in the inequality
(\ref{uncert3}), we get

\be \D x \D p \geq \frac{\hbar}{2} \le[1+ \frac{5}{3}~
(1+\mu)~\b_0~ \ell_p^2~ \frac{\D p^2}{\hbar^2} \ri]\,.
\la{ineqI} \ee

By solving the inequality (\ref{ineqI}) as quadratic equation
in $\D p$, we obtain

\be \frac{\D p}{\hbar}\geq\frac{\D x}{\frac{5}{3}~(1+\mu)~\b_0
\ell_{p}^2}\le[1- \sqrt{1-\frac{\frac{5}{3}~(1+\mu)~
\b_0\ell_{p}^2} { \D x^2}}~~\ri] \,.\la{gupso} \ee

Where we considered only the negative sign$(-)$ solution which
gives the standard uncertainty relation as $\frac{\ell_p}{\D
x}\rightarrow 0$.

Using the same  arguments that were used in Sec. $2$,  the
modified Hawking temperature will be given by:

\bea
T_H^{\prime}&=& \frac{D-3}{\pi\b_0} \frac{M_p c^2}{\frac{\frac{5}{3}~(1+\mu)}{2}}m^{\frac{1}{D-3}}\l_D
\le[1-\sqrt{1-\frac{\frac{5}{3}~(1+\mu)~ \b_0 }{4 \l_D^2 m^{\frac{2}{D-3}}}}~~\ri].\la{modT}~~\\
&=&2 T_H \le[1+\sqrt{1-\frac{\frac{5}{3}~(1+\mu)~~\b_0 }{4\l_D^2 m^{\frac{2}{D-3}}}}~~\ri]^{-1}
\eea

The modified Hawking temperature is physical as far as the
black hole mass satisfies the following inequality:

\be
4~\l_D^2 ~m^{\frac{2}{D-3}} \geq \frac{5}{3}~(1+\mu)~ ~\b_0
\ee

This tells us the black hole should have minimum mass $M_{min}$
given by

\be
M_{min}=M_p \le(\sqrt{\frac{\frac{5}{3}~(1+\mu)~}{4}}~~\ri)^{D-3} \frac{D-2}{8~ \Gamma({\frac{D-1}{2}})}
(\sqrt{\b_0}~\sqrt{\pi})^{D-3}. \la{MinM}
\ee

Here we note the minimum mass of the BH is different from the
one obtained in \cite{Cavaglia:2003qk}. There is a new factor
$(\sqrt{5(1+\mu)/12~~})^{D-3}$
which would give higher values for the minimum mass of the BHs.
This factor appeared because we considered the GUP that follows
from Jacobi identity (see \ref{uncert1}) \cite{kempf}
and which Cavaglia et al.\cite{Cavaglia:2003qk} did not consider.

The endpoint of Hawking evaporation in the GUP-case is
characterized by a Planck-size remnant with maximum temperature
\be T_{max}~=~2 ~T_H. \ee

The emission rate can be calculated using Stefan-Boltzmann Law,
using Eq.\ ({\ref{dMdt},\ref{rate1}}), Since the BH mostly
emitting on the brane, so we consider 4-dimensional brane, so
we get

\be
\frac{dm}{dt}= - 16  \frac{\mu^{\prime}}{t_p} m^{\frac{-2}{D-3}}
\le[1+\sqrt{1-\frac{\frac{5}{3}~(1+\mu)~\b_0 }{4\l_D^2 m^{\frac{2}{D-3}}}}~~\ri]^{-4}
\ee

The entropy can be calculated from the first law of
BH-thermodynamics,

\be dS=\frac{2~\pi}{D-3} \l_D~
m^{\frac{1}{D-3}}~\le[1+\sqrt{1-\frac{\frac{5}{3}~(1+\mu)~\b_0
}{4\l_D^2 m^{\frac{2}{D-3}}}}~~\ri]dm. \ee

The specific heat has been calculated in GUP-case to give

\bea {\cal C}\equiv T\frac{\partial S}{\partial T}= M_p c^2
\frac{\partial m}{\partial T}=-2\pi\l_d
m^{(d-2)/(d-3)}\sqrt{1-\frac{\frac{5}{3}~(1+\mu)~\b_0 }{4\l_D^2
m^{\frac{2}{D-3}}}} \left(1+
\sqrt{1-\frac{\frac{5}{3}~(1+\mu)~\b_0 }{4\l_D^2
m^{\frac{2}{D-3}}}}~\right) \la{sh}.~~~ \eea

We note the BH specific heat vanishes at the minimum BH-mass.
Therefore, the BH cannot exchange heat with the surrounding
space. This may solve the problem of \emph{\textbf{catastrophic
evaporation}} of the BH that was discussed in the previous
section.

\section{GUP linear and quadratic in $\D p$ and BH thermodynamics }



In this section, we would like to find the corresponding
 inequality for Eq.\ (\ref{comm01}) in $(D-1)$-dimensions. Eq.\
 (\ref{comm01}) gives with using the argument used in
 \cite{Cavaglia:2003qk}, \be \D x \D p \geq
 \frac{\hbar}{2}\le[1- \a \langle p \rangle- \a \langle
 \frac{p_i^2}{p} \rangle
+\a^2 \langle p^2 \rangle + 3 \a^2 \langle p_i^2\rangle \ri]\,.
\la{ineq} \ee

Using arguments in the Sec. 3, Eqs.\ (\ref{3n},\ref{arg}), in
the inequality (\ref{ineq}), we get

\be \D x \D p \geq \frac{\hbar}{2} \le[1- \a_0 ~\ell_p~
\le(\frac{4}{3}\ri)~\sqrt{\mu}~~ \frac{\D p}{\hbar}+ ~2~
(1+\mu)~ \a_0^2 ~\ell_p^2 ~ \frac{\D p^2}{\hbar^2} \ri]\,.
\la{ineqII} \ee The last inequality is ( and as far as we know
the only one) following from Eq.\ (\ref{comm01}).

By solving the inequality (\ref{ineqII}) as quadratic equation
in $\D p$, we obtain \be \frac{\D p}{\hbar}\geq\frac{2 \D
x+\a_0
~\ell_p~\le(\frac{4}{3}~\sqrt{\mu}~\ri)}{4~(1+\mu)~\a_0^2~\ell_{p}^2}\le[1-
\sqrt{1-\frac{8~(1+\mu)~\a_0^2\ell_{p}^2} {\le(2 \D x+\a_0
\ell_p\le(\frac{4}{3}\ri) ~\sqrt{\mu}~\ri)^2}}~\ri]
\,.\la{gupso} \ee

Where we considered only the negative sign$(-)$ solution which
gives the standard uncertainty relation as $\frac{\ell_p}{\D
x}\rightarrow 0$.

Using the same  arguments that were used in Sec. $2$, the
modified Hawking temperature will be given by:

\bea T_H^{\prime}&=& \frac{D-3}{\pi\a_0^2} \frac{M_p
c^2}{(1+\mu)}\le(m^{\frac{1}{D-3}}\l_D + \frac{\a_0~
\sqrt{\mu}}{3}\ri)
\le[1-\sqrt{1-\frac{(1+\mu)~\a_0^2 }{2 \le(\l_D m^{\frac{1}{D-3}}+\frac{\a_0\sqrt{\mu}}{3}\ri)^2}}\ri]\la{modT}~~~~~~~~\\
&=&2 T_H \le(1+\frac{\a_0 ~\sqrt{\mu}~}{3~\l_D
m^{\frac{1}{D-3}}}\ri)^{-1} \le[1+\sqrt{1-\frac{(1+\mu)~ \a_0^2
}{2 \le(\l_D m^{\frac{1}{D-3}}+ \frac{\a_0 ~\sqrt{\mu}}{3}
\ri)^2}}\ri]^{-1} \eea

The modified Hawking temperature is physical as far as the
black hole mass satisfies the following inequality:

\be (1+\mu)~\a_0^2  \leq  2 \le(\l_D
m^{\frac{1}{D-3}}+\frac{\a_0~ \sqrt{\mu}}{3}~\ri)^2 \ee This
tells us the black hole should have minimum mass $M_{min}$
given by

\be
M_{min}=M_p \le(\sqrt{\frac{(1+\mu)}{2}}-\sqrt{\frac{\mu}{9}}\ri)^{D-3} \frac{D-2}{8 \Gamma({\frac{D-1}{2}})}
(\a_0 \sqrt{\pi})^{D-3}. \la{MinMI}
\ee

Here we note the minimum mass of the BH is different from the
one obtained in \cite{Cavaglia:2003qk}. There is a new factor
$\le(\sqrt{(1+\mu)/2}-\sqrt{\mu/9}\ri)^{D-3}$
which would give higher values for the minimum mass of the BHs.
This factor appeared because we considered our proposed GUP in \cite{advplb,Ali:2010yn,Das:2010zf} that follows
from Jacobi Identity (see \ref{comm01}) \cite{advplb,Ali:2010yn,Das:2010zf}.

The endpoint of Hawking evaporation in the GUP-case is
characterized by a Planck-size remnant with maximum temperature
\bea T_{max} \approx 2
\le[\frac{\frac{3(1+\mu)}{2}+\sqrt{\frac{\mu(\mu+1)}{2}}}{\frac{3}{2}
+ \frac{7}{6} \mu}\ri]~ T_H. \eea

The emission rate can be calculated using Stefan-Boltzmann Law,
using Eq.\ ({\ref{dMdt},\ref{rate1}}), we get for 4-dimensional
brane:

\be
\frac{dm}{dt}= -16  \frac{\mu^{\prime}}{t_p} m^{\frac{-2}{D-3}} \le(1+\frac{\a_0~\sqrt{\mu}}{3\l_D m^{\frac{1}{D-3}}}\ri)^{-4}
\le[1+\sqrt{1-\frac{(1+\mu) \a_0^2 }{2 \le(\l_D m^{\frac{1}{D-3}}+\frac{\a_0\sqrt{\mu}}{3}\ri)^2}}\ri]^{-4}
\ee

The entropy can be calculated from the first law of BH-thermodynamics,
\be
dS=\frac{2\pi}{D-3} \l_D m^{\frac{1}{D-3}}\le(1+\frac{\a_0~\sqrt{\mu}}{3\l_D m^{\frac{1}{D-3}}}\ri)
\le[1+\sqrt{1-\frac{(1+\mu) \a_0^2 }{2 \le(\l_D m^{\frac{1}{D-3}}+\frac{\a_0\sqrt{\mu}}{3}\ri)^2}}\ri]~dm.
\ee

The specific heat has been calculated in GUP-case to give

\bea
{\cal{C}}&=&-\frac{2 \pi}{\l_D} m^{\frac{D-4}{D-3}} \le(\l_D m^{\frac{1}{D-3}}+ \frac{\a_0~\sqrt{\mu}}{3} \ri)^2
\sqrt{1-\frac{(1+\mu) \a_0^2 }{2 \le(\l_D m^{\frac{1}{D-3}}+\frac{\a_0\sqrt{\mu}}{3}\ri)^2}}\nn\\&&
\le[1+\sqrt{1-\frac{(1+\mu) \a_0^2 }{2 \le(\l_D m^{\frac{1}{D-3}}+\frac{\a_0\sqrt{\mu}}{3}\ri)^2}}\ri].
\eea
We note the BH specific heat vanishes at the minimum BH-mass. Therefore, the BH cannot exchange heat with the surrounding space.

\section{NO black holes at LHC current energy scales due to GUP}

In this section, we use the calculations in Sec. 3, and Sec. 4
to investigate whether black holes could be formed at  LHC
energy scales . From Eqs.\ (~\ref{MinM},~\ref{MinMI}~), we note
that black holes can be formed with masses larger than $M_p$ in
$D$-dimensions. The model of GUP- black holes in higher
dimensions has three unknown parameters: $D$, $M_P$, and
$\b_0$($ \a_0$). If we fix the GUP-parameters to be
$\b_0=1$($\a_0=1$). In this case, the values for the minimum
black hole masses in the extra dimensions, using Eqs.\
(\ref{MinM},\ref{MinMI}), are shown in the following Table.


\begin{table*}[h]
\caption{BH minimal mass for different dimensions using the latest observed limits on the ADD model parameter $M_p$ in
\cite{Chatrchyan:2012pa}. }
\begin{center}
\begin{tabular}{|c|c|c|c|c|}\hline
 & &\footnotesize{GUP-Quadratic:$\b_0=1$}& \footnotesize{GUP-Linear$\&$Quadratic:$\a_0=1$}
 &\footnotesize{GUP-Quadratic of}\cite{Cavaglia:2003qk}\\ \hline\hline
$D$ &$M_p$&$ M_\mathrm{min}~~$& $M_\mathrm{min}~$&$M_\mathrm{min}~$~\\\hline
6 & $4.54$ ~TeV&$>60.7$ ~TeV& $>13.8 $ ~TeV &$>2.1~$ TeV \\\hline
7 &$3.51$ ~TeV&$>362$ ~TeV& $> 46.3$ ~TeV &$>3.1$~ TeV \\\hline
8 &$2.98$ ~TeV &$>2714$ ~TeV& $>196$ ~TeV &$>3.9$~TeV \\\hline
9 & $2.71$ ~TeV&$>24 \times 10^3$ ~TeV& $>982$ ~TeV &$>4.5$~TeV\\\hline
10 & $2.51$ ~TeV& $>2.3 \times 10^5$ ~TeV& $>5.5\times10^3$ ~TeV & $>4.7$~ TeV \\\hline
\end{tabular}
\end{center}
\label{numbertable2}
\end{table*}

\begin{table*}[h]
\caption{The Schwarzschild radius $R_s=
\frac{1}{\sqrt{\pi}M_p}\le[ \frac{M_{BH}}{M_p} \frac{8
\Gamma(\frac{D-1}{2})}{D-2}\ri]^{\frac{1}{D-3}}$ using the latest observed limits on the ADD model parameter $M_p$ in
\cite{Chatrchyan:2012pa}.}
\begin{center}
\begin{tabular}{|c|c|c|c|}\hline
 & \footnotesize{~GUP-Quadratic:$\b_0=1$}& \footnotesize{GUP-Linear$\&$Quadratic:$\a_0=1$}
 &\footnotesize{GUP-Quadratic of\cite{Cavaglia:2003qk}}\\ \hline\hline
$D$ &$ R_s~~$& $R_s~$&$R_s~$~\\\hline
6  & $>0.41$ & $ >0.25 $  &$>1 ~$  \\\hline
7 & $>0.69$ & $ >0.41 $  &$>0.99$ \\\hline
8 &$>0.99$ & $> 0.58 $  &$>0.99$\\\hline
9 & $>1.31$ & $ >0.77 $  &$>1$\\\hline
10 & $>1.64$ & $>0.96$ & $>0.99$ \\\hline
\end{tabular}
\end{center}
\label{numbertable3}
\end{table*}

In table~\ref{numbertable2}, a BH in $D$-dimensions at fixed
$\b_0=1$ can form only for energies equal to or larger than its
minimum mass. We consider the latest observed limits on the ADD model parameter $M_p$ in
\cite{Chatrchyan:2012pa}.


This means BH's ( If GUP-quadratic in $\D p$ is only
considered) in $D=6$ can form only at energies not less than
$60.7$~TeV, and for $D= 8$, they can form only for energies
not less than $2714$~TeV,
and for BH's in $D=10$, they can only form for energies not less than $2.3\times 10^5$~TeV.\\
Turning to  GUP-linear and quadratic in $\D p$  case, we found
that the black hole can be formed at energies less than the
ones predicted by GUP-quadratic case, but they are still larger
than the current energy scales of LHC. The BH's in $D=6$ can
form only at energies not less than $13.8$~TeV, and BH's can
form in $D=7$ for energies not less than
$46.3$ TeV.\\

In the table ~\ref{numbertable2},  we compare  our results with
the results proposed in \cite{Cavaglia:2003qk}. The previous
studies in \cite{Cavaglia:2003qk} predicted that BH's might be
seen at the energy scales of LHC in disagreement with the
recent experimental results of LHC \cite{EXPERIMENT,EXPERIMENT2}. Our
results possibly agrees with the results of the experiment
\cite{EXPERIMENT, EXPERIMENT2}. We
found that black holes can be formed at energies much higher
than the current energy scales of LHC. Predictions of mini
black holes forming at collision energies of a few TeV's were
based on theories that consider the gravitational effects of
extra dimensions of space like string theory and large extra
dimensions theories
\cite{ArkaniHamed:1998rs,Antoniadis:1998ig,Randall:1999ee,Randall:1999vf,Dimopoulos:2001hw}.
But scientists at the Compact Muon Solenoid (CMS) detector in
LHC  are excluding semiclassical and quantum black holes with masses below $3.8$ to
$5.3$ TeV. Our proposed model of GUP can possibly justify why
higher energies larger than the current scale of LHC is needed
to form mini black holes.\\

\section{Conclusions}

We investigated whether the GUP can
explain the formation of black holes at energies higher than
the energy scales of LHC to explain the recent experimental results that
were obtained at LHC \cite{EXPERIMENT, EXPERIMENT2}. We have shown that, by studying
Hawking-Uncertainty connection, the black holes can be formed
in the range between $13.8- 5.5\times 10^3$~TeV for GUP-linear and
quadratic in $\D p$ for values of $D$ between $6$ and $10$, and they can be formed in the range
between $60-2.3\times 10^5$~TeV for GUP-quadratic in $\D
p$ for values of $D$ between $6$ and $10$. Both cases say black holes can be formed at energies higher
than the current energy scales of LHC. We conclude that
mechanisms such as GUP may be necessary to explain the recent experimental results. In the
future, it would be appropriate to apply our approach on the
calculations of the cosmological constant, black body
radiation, etc. We hope to report on these in the future.
\\

{\large{\bf Acknowledgments} ; }\\ The author gratefully thanks Saurya Das for many enlightening
discussions and comments to finish this paper. The author gratefully thanks the anonymous referee
for useful comments and suggestions which helped to improve the paper.


\end{document}